\documentclass[aps, prb, twocolumn, showpacs, amssymb, superscriptaddress, groupedaddress]{revtex4}
\usepackage{bm}
\usepackage{commath}
\usepackage{graphicx}
\usepackage{setspace}
\usepackage{amsmath}
\usepackage{amssymb}
\usepackage{physics}
\usepackage[normalem]{ulem}
\usepackage{color}

\begin{document}

\title{Chiral Quasiparticle Tunneling Between Quantum Hall Edges in Proximity with a Superconductor}
\author{M. T.~Wei} \affiliation{Department of Physics, Duke University, Durham, NC 27708, USA}
\author{A. W.~Draelos} \affiliation{Department of Physics, Duke University, Durham, NC 27708, USA}
\author{A.~Seredinski} \affiliation{Department of Physics, Duke University, Durham, NC 27708, USA}
\author{C. T.~Ke} \affiliation{Department of Physics, Duke University, Durham, NC 27708, USA}
\author{H.~Li} \affiliation{Department of Physics and Astronomy, Appalachian State University, Boone, NC 28607, USA}
\author{Y.~Mehta} \affiliation{Department of Physics and Astronomy, Appalachian State University, Boone, NC 28607, USA}
\author{K.~Watanabe} \affiliation{Advanced Materials Laboratory, NIMS, Tsukuba 305-0044, Japan}
\author{T.~Taniguchi} \affiliation{Advanced Materials Laboratory, NIMS, Tsukuba 305-0044, Japan}
\author{M.~Yamamoto}  \affiliation{Center for Emergent Matter Science (CEMS), RIKEN, Wako-shi, Saitama 351-0198, Japan}
\author{S.~Tarucha}  \affiliation{Center for Emergent Matter Science (CEMS), RIKEN, Wako-shi, Saitama 351-0198, Japan}
\author{G.~Finkelstein} \affiliation{Department of Physics, Duke University, Durham, NC 27708, USA}
\author{F.~Amet} \affiliation{Department of Physics and Astronomy, Appalachian State University, Boone, NC 28607, USA}
\author{I. V. Borzenets}\email[]{iborzene@cityu.edu.hk} \affiliation{Department of Physics, City University of Hong Kong, Kowloon, Hong Kong SAR}

\date{\today}

\begin{abstract}

We study a two-terminal graphene Josephson junction with contacts shaped to form a narrow constriction, less than $100$nm in length. The contacts are made from type II superconducting contacts and able to withstand magnetic fields high enough to reach the quantum Hall (QH) regime in graphene. In this regime, the device conductance is determined by edge states, plus the contribution from the constricted region. In particular, the constriction area can support supercurrents up to fields of $\sim 2.5$T. Moreover, enhanced conductance is observed through a wide range of magnetic fields and gate voltages. This additional conductance and the appearance of supercurrent is attributed to the tunneling between counter-propagating quantum Hall edge states along opposite superconducting contacts.

\end{abstract}


\maketitle

In the past few years, there has been a renewed interest in quantum Hall (QH) states supported along superconducting (SC) materials. Experimentally, this was prompted by several groups successfully making high-transparency type II superconducting contacts to both encapsulated graphene and III-V semiconductor heterostructures \cite{BenShalom2016, Calado2015, Wan2015, Amet2016, Draelos2018, Seredinski2018, Seredinski2019, Lee2017, Park2017, Sahu2018}. Meanwhile, theoretical works have predicted multiple exciting phenomena in structures combining the quantum Hall effect and superconductivity \cite{Ma1993, Zyuzin1994, Fisher1994, Takagaki1998, Hoppe2000, Chtchelkatchev2007, Khaymovich2010, vanOstaay2011,  Clarke2013, Mong2014, Alicea2016, Clarke2014, SanJose2015, Zocher2016, Cohnitz2017, Alavirad2017, Gamayun2017, Huang2017, Finocchiaro2018, Repellin2018}. In particular, it is expected that Andreev edge states (AES) -- hybrid modes involving a linear superposition of electron and hole states -- should be formed at these QH-SC interfaces \cite{Park2017, Hoppe2000, Takagaki1998, Chtchelkatchev2007, Khaymovich2010}. Furthermore, these structures have been predicted to support Majorana zero modes and parafermions when the symmetry-breaking QH edge states are coupled to SC \cite{Clarke2013,Clarke2014,Mong2014,Alicea2016}. Here we explore AES and tunneling between two superconducting contacts across a narrow region of graphene in the quantum Hall regime. 

\begin{figure}[tbp]
\begin{center}
\includegraphics[width=\linewidth]{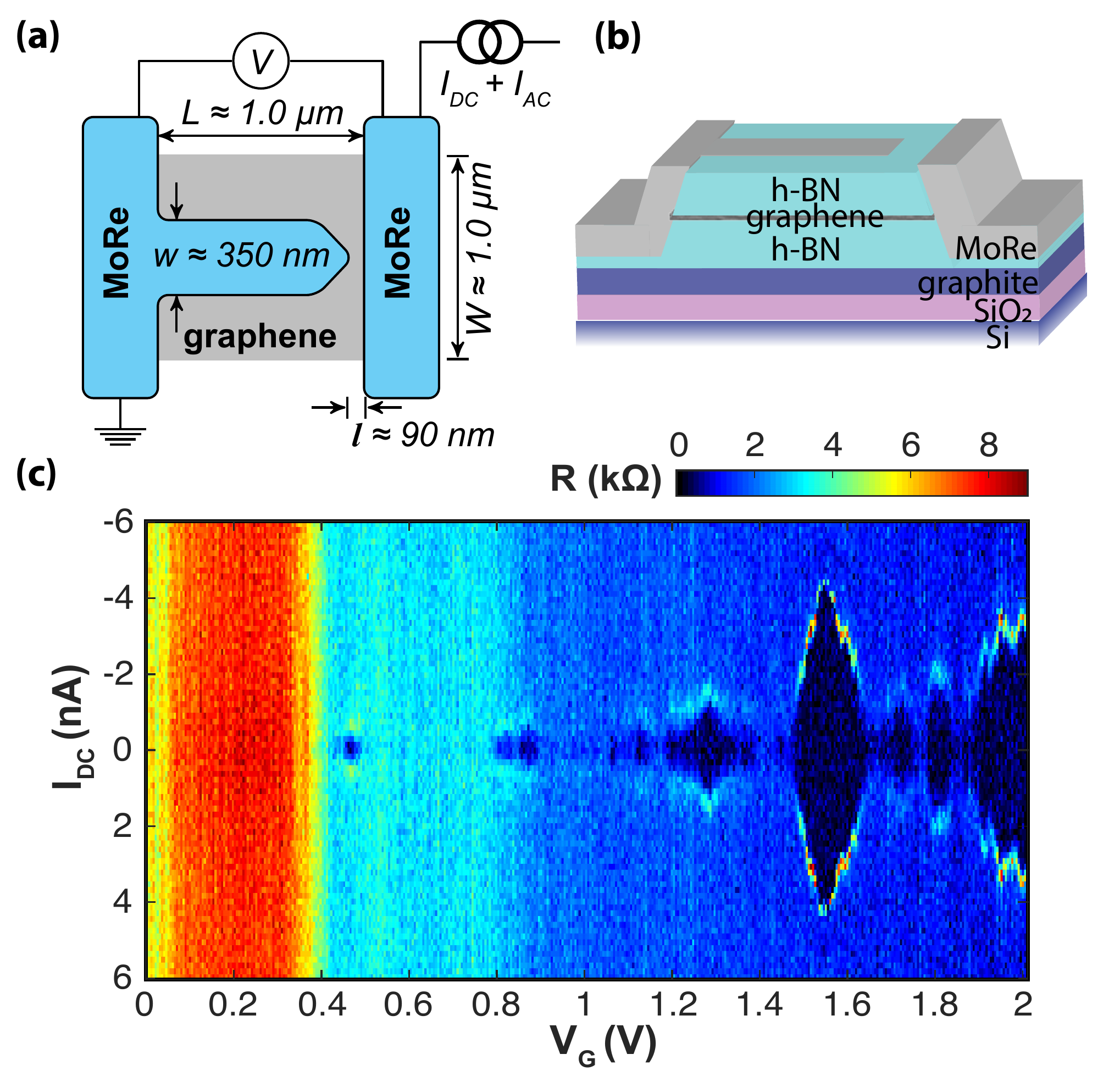}
\end{center}
	\caption{(a) Schematic of the graphene Josephson junction with asymmetric contacts, measured with a four-probe current-biased setup. The finger of the T-shaped contact is separated from the opposing contact by about $90$nm. (b) 3D representation of the T-shaped junction illustrating the graphene-hBN stack with the graphite back gate. (c) Differential resistance $R=dV/dI$ versus DC bias current $I_{DC}$ and gate voltage $V_G$ taken at a perpendicular magnetic field of $B=2$T. Pockets of suppressed resistance (superconductivity) are observed on the clear quantum Hall plateaus. 
	}
\label{fig:Fig1} 
\end{figure}

Our device design is shown in Figure ~\ref{fig:Fig1}. A graphene crystal of $1\mu$m$\times 1\mu$m is contacted on two sides by the superconductor molybdenum rhenium (MoRe). The contacts are  asymmetric, with one interface being flat, and the other having a ``T'' shape. The 350 nm-wide leg of the T extends into the graphene, such that the shortest separation between contacts is $l\sim 90$nm. The graphene device is assembled by a standard stamping technique \cite{Dean2010}, where monolayer graphene is sandwiched by hexagonal boron nitride (hBN) and placed onto a graphite back gate (Fig. \ref{fig:Fig1}(b)). The back gate-graphene distance is $\sim 40$nm (confirmed by atomic force microscopy), and metal leads of Cr/Au ($5$nm/$110$nm) are used to make contact with the back gate. Carefully calibrated etching of the stack allows us to avoid shorting to the graphite back gate.

The sample was measured with a pseudo four-probe setup in a Leiden Cryogenics dilution refrigerator with a base electron temperature of $\sim 50$mK (at zero field) to $\sim 60$mK (at high fields). A DC bias current along with a small AC excitation is supplied by a combination of an NI USB-6363 digital acquisition device and a lock-in amplifier. The measured voltage is initially amplified by a home-made, low-frequency, low-noise amplifier. Three-stage RC filtering, stainless steel powder filter, and resistive lines were all employed to lower the high frequency noise that can suppress the supercurrent.  Carrier density in the graphene was tuned via a back gate voltage applied to the graphite layer, where the gate capacitance is $C_G \approx 70$nF/cm$^2$. Magnetic fields are applied perpendicular to the plane of the graphene sheet. 

\begin{figure}[tbp]
	\center
	\includegraphics[width=\linewidth]
	 {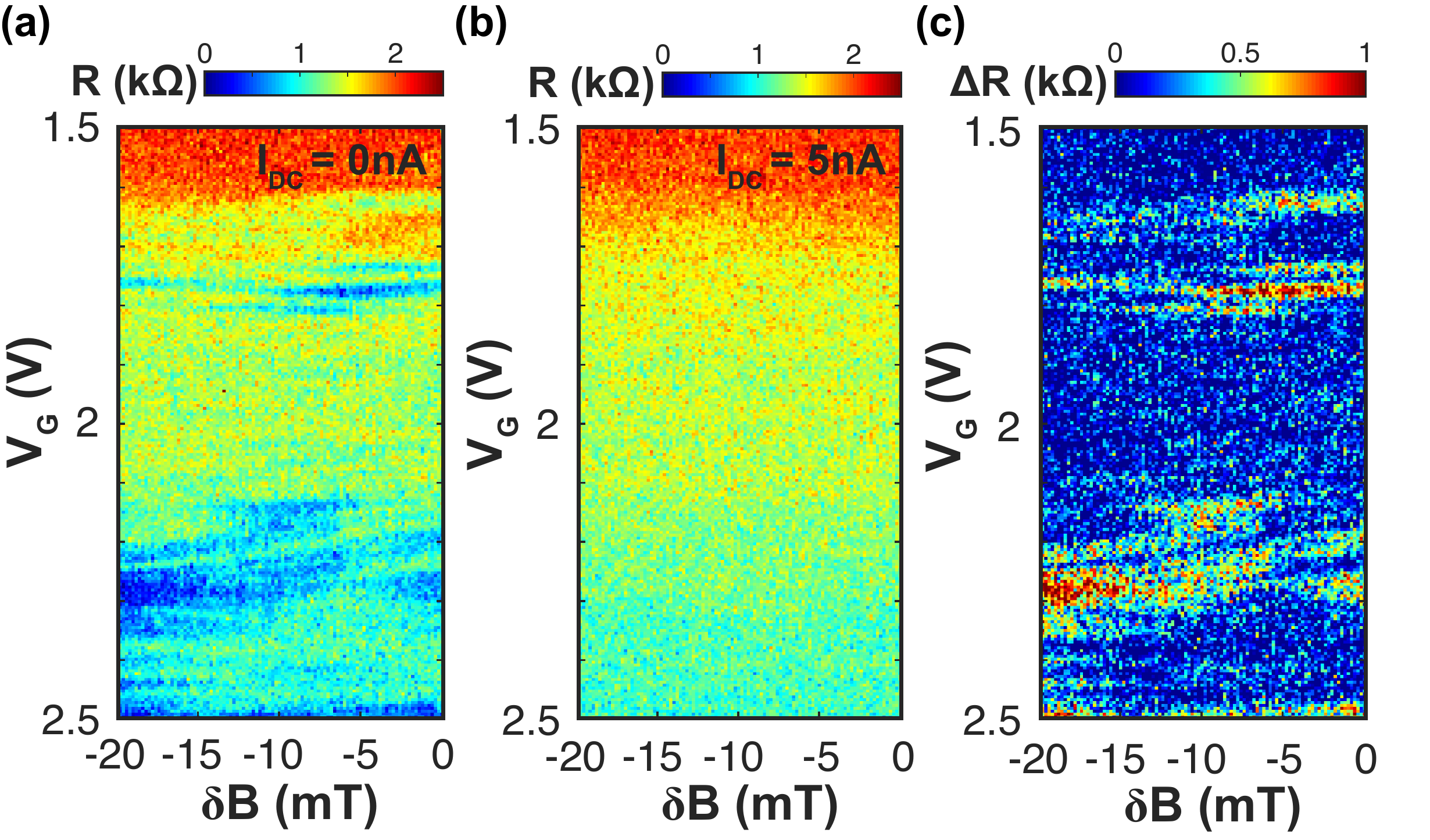}
	\caption[Supercurrents at 2.5T]{ The dependence of differential resistance $R$ on magnetic field: $B=2.5$T$+\delta B$ as a function of gate voltages $V_G$ and bias current. (a) Resistance dips at $I_{DC}=0$nA indicate pockets of supercurrent. (b) Superconducting signatures are fully suppressed at $I_{DC}=5$nA. (c) The resistance difference $\Delta R$ between $I_{DC}=5$nA and $0$nA. No periodic oscillations of supercurrent in field are observed. This suggests that the supercurrent is not mediated by the QH states along the graphene edges, as a SQUID-like pattern would be expected to emerge.} 
\label{fig:Fig2}
\end{figure}


The differential resistance $R=dV/dI$ is shown in Fig. \ref{fig:Fig1}(c) as a function of the DC bias current $I_{DC}$ and back gate voltage $V_G$, taken at magnetic field $B= 2$T. Near zero bias, areas of suppressed resistance can be clearly observed, indicating the presence of supercurrent. Pockets of supercurrent are seen at multiple locations in $V_G$, both on and off the conductance plateaus. 
Fig. \ref{fig:Fig2}(a) shows this same differential resistance vs $\delta B$ and $V_G$ at zero bias, as the magnetic field is varied only slightly to investigate the periodicity of the observed supercurrent\cite{Amet2016, Draelos2018,vanOstaay2011}.  
Low resistances at $I_{DC}=0$ again show the supercurrent, which is contrasted by Fig. \ref{fig:Fig2}(b) where the supercurrent is fully suppressed due to a DC bias current of $I_{DC}=5$nA. 
The suppression of resistance can be quantified by subtracting the zero bias resistance from the high bias resistance $\Delta R=R_{5nA}-R_{0nA}$, shown in Fig. \ref{fig:Fig2}(c), where high $\Delta R$ indicate regions of supercurrent\cite{ Draelos2018}. 

Unlike previous works, these pockets of supercurrent do not show periodic oscillations with magnetic field \cite{Amet2016, Draelos2018}. For a Josephson junction of area $A\approx 0.7\mu$m$^2$ with supercurrent supported along the circumference, oscillations with a period of $\Delta B\approx 0.5$mT are expected \cite{Amet2016, Draelos2018,Tinkham}. Instead, the observed features are aperiodic and change on the scale of $\Delta B\sim 10$mT, suggesting that the supercurrent does not flow along the graphene-vacuum edges. Moreover,  the measured normal resistance of the QH plateaus is lower than the expected quantized fractions of $h/2e^2$. This strongly suggests the existence of additional conducting channels beyond the standard QH edge states. 

\begin{figure}[ht]
\includegraphics[width=\linewidth]
	{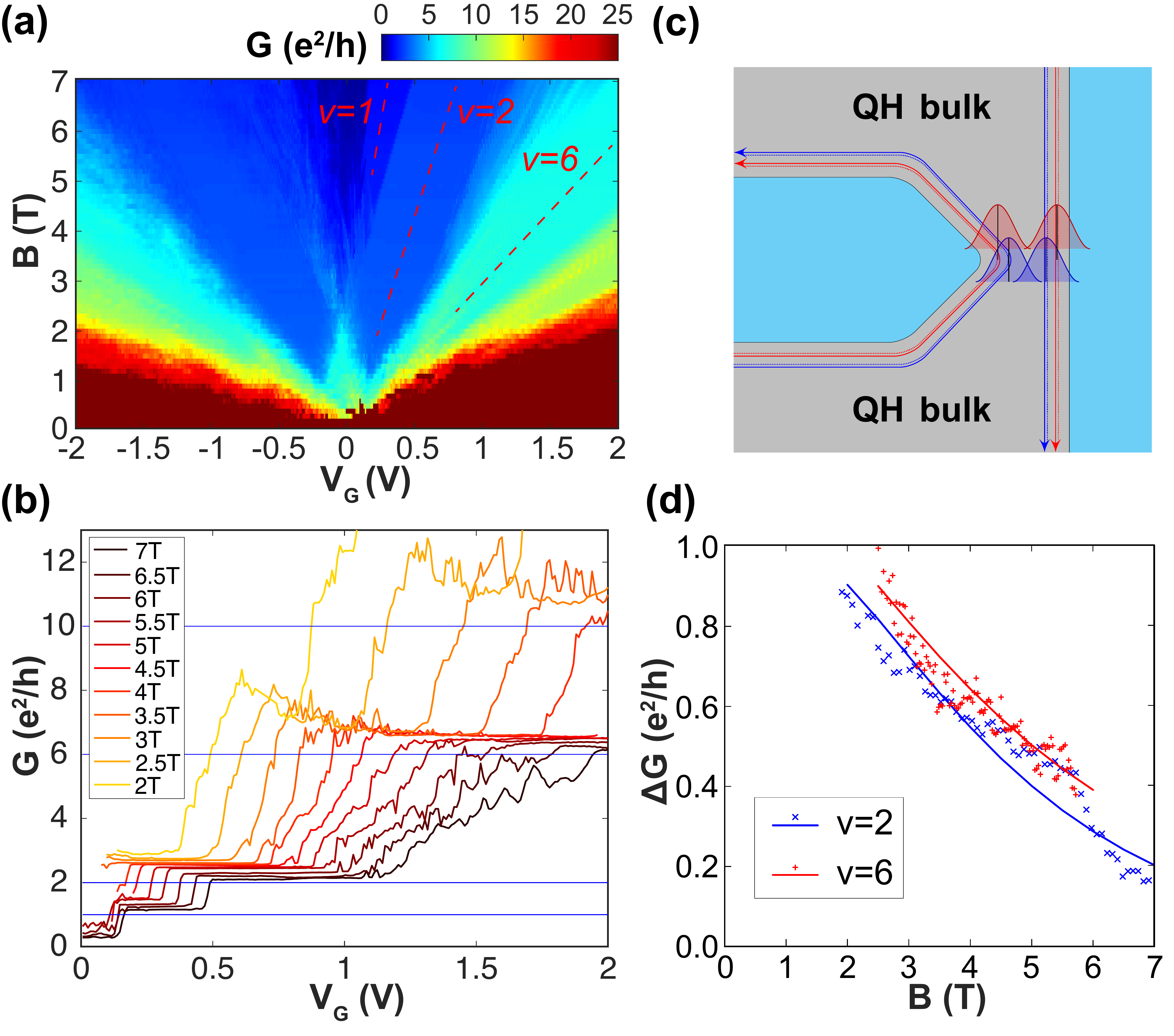}
	\caption  [Field-dependent conductance]{Field-dependent conductance. (a) The fan diagram of measured conductance from 0 to 7 T. The red dashed lines follow the centers of the $\nu=1,2,6$ plateaus. 
		(b) The measured conductance as a function of back gate voltages $V_G$ taken at several magnetic fields from $2$ to $7$ T. (c) The probability amplitudes of the wavefunctions on each side are illustrated by the red and blue peaks for the $\nu=2$ and $6$ states, respectively. The tunneling can be explained by the overlap between these wavefunctions in all chiral channels, where the states with higher filling factors show greater overlapping. (d) Enhancement in the measured conductance $\Delta G$ for the marked line cuts in (a) at $\nu =2$,$6$ along the center of the QH plateaus. The dots are the measured data, while the solid lines are fitted conductances simulating the contribution due to tunneling via overlapping QH states running along the graphene-superconductor interface. }
\label{fig:Fig3}
\end{figure}

We next measure the sample conductance using only a DC bias ($I_{DC}=5$nA) to avoid measurement errors due to stray capacitance. Figure \ref{fig:Fig3}(a) shows the fan diagram of conductance vs back gate voltage and magnetic field up to $7$T. Above $4$T, we see the $\nu=1$ plateau developing in addition to the $\nu=2, 6, 10$ plateaus previously studied. 
Fig. \ref{fig:Fig3}(b) shows selected cross sections of the conductance as a function of back gate voltage at magnetic fields from $2$T to $7$T,  compared to the expected value of each plateau (flat blue lines). It is apparent that the height of each plateau decays monotonically with increasing magnetic field, though without fully reaching the expected value of QH conductance. Nevertheless, each plateau is flat with respect to $V_G$, indicating that the additional conductance we observe is not due to bulk density changes. Note that this decreasing conductance with increasing field cannot be attributed to growing finite resistance of the superconductor near its critical field (as in the case of niobium in Ref. \onlinecite{Rickhaus2012}) because MoRe alloys do not exhibit a finite resistance for the magnetic fields used here\cite{Calado2015}.


The existence of non-periodic supercurrents at 2.5T and field-dependent conductance can both be attributed to the coupling of QH edge states across the short $90$nm channel. For lower magnetic fields, when the cyclotron radius $r=\hbar \sqrt{n\pi}/eB>l/2 \approx 45$nm, the short channel is in the semiclassical regime\cite{BenShalom2016}. (Here $n$ is the carrier density). As such, supercurrent can be supported by conventional ABSs. When $r\ll 45$nm, the supercurrents could be mediated via quantum mechanical tunneling between QH edge modes \cite{Moon93, Kim03, Roddaro03, marcus, martinis2013}. Hence, the enhanced conductance $\Delta G$ approaching $\nu e^2/h$ suggests that the overall conductance can be written as 
\begin{equation*}
G_{\mathrm{total}}=G_{\mathrm{QH}}+G_{\mathrm{tunneling}},
\end{equation*}
where $G_{\mathrm{QH}}$ is the expected QH conductance of the edge channels along the vacuum edges and $G_{\mathrm{tunneling}}$ is the additional conductance from the T-shaped short channel. 

Demonstrated schematically in Fig. \ref{fig:Fig3}(c), the red and blue solid/dashed lines represent the counter-propagating chiral electron-hole hybrid modes of $\nu=2$ and $6$, respectively. These can individually be described by parabolic cylindrical wavefunctions \cite{Hoppe2000} at zero bias current. These wavefunctions are intrinsically different from the wavefunctions of the regular QH edge states\cite{Zheng2002}. They are centered at distance $X_0=k_xl_B^2\propto V_G/B$ away from a superconductor-graphene edge, where $k_x$ is the wavevector parallel to the contacts and $l_B$ is the magnetic wavelength $\approx 26$nm$/\sqrt{B}$. Moreover, they have a characteristic width that also scales with magnetic field as $W_0\propto l_B\propto1/\sqrt{B}$. The overlap between these wavefunctions propagating along each contact in the short channel is what mediates the supercurrent by tunneling.

Qualitatively, as the field increases, the position of the center of an edge state at each contact will move toward the contact (or move away from each other). Additionally, the width of an edge state wavefunction will decrease with $B$. Thus, as $B$ increases, the overlap between edge state wavefunctions at the constriction decreases. This reduces the tunneling and decreases $G_{\mathrm{tunneling}}$. However, for a short enough channel, the overlap between two chiral edge states near both contacts will contribute a nonzero conductance even at $7$T.  

We estimate the enhancement in conductance $\Delta G$ due to tunneling at the constriction by simulating the amount of physical overlap of the two edge-state wavefunctions: 
\begin{equation*}
\Delta G= A \int_{-\infty}^{	\infty} \Psi_L(\frac{x-X_0}{W_0})\Psi_R(\frac{l+X_0-x}{W_0}) dx
\end{equation*}

Here, $\Psi_L$ and $\Psi_R$ represent the wavefunctions of the left and right QH edge states whose position and width are determined by $V_G$ and $B$ via $X_0$ and $W_0$; $l=90$nm is the length of our constriction; and $A$ is a proportionality constant determined in part by the constriction width. Note that such scheme of conduction is similar to tunneling \emph{across} a point contact constriction in the quantum Hall regime \cite{Moon93, Kim03, Roddaro03, marcus, martinis2013}. Here, however, the constriction is defined by the gapped QH bulk on both sides, instead of the point contact split gates.

For simplicity, we vary $V_G$ such that the wavefunctions center offset $X_0$ remains constant with respect to increasing $B$; thus, only $W_0$ is changed. This is visually represented as the red dashed lines in Figure \ref{fig:Fig3}(a). The measured enhanced conductance $\Delta G$ along those lines for $\nu=2$ and $\nu=6$ is shown in Figure \ref{fig:Fig3}(d) as dots. Fits of our simulated $\Delta G$, with $A$ and $X_0$ as the fitting parameters, are plotted as lines. The wavefunctions $\Psi_L$ and $\Psi_R$ were roughly estimated as Gaussians with degeneracy 2 for $\nu=2$ and 4 for $\nu=6$ (as the edge states with higher filling factors are placed closer together). We find that even such a rough estimate produces an acceptable fit to our data (with a reasonable fitted $X_0\sim 20-30$nm). In fact, taking a more complicated approximation to $\Psi_L$ and $\Psi_R$ (following Ref. \onlinecite{Zheng2002}) did not result in significant improvements to the fit.

In our device, we expect the conductances of the QH edge states to reach their theoretically expected values by $B\approx 10$T. Knowing both the amount of overlap and the strength of interactions between two QH edge states is important when coupling them to produce topological states such as paraferimons \cite{Alicea2016}. This experiment provides an important first step towards the design of QH-SC structures that are capable of supporting such non-Abelian excitations.

In conclusion, a short channel in a Josephson junction with T-shaped asymmetric contacts has been shown to mediate a non-periodic supercurrent and cause a nontrivial extra conductance that gradually decays at higher fields. This result is the first tunneling evidence of the chiral electron-hole hybrid modes between two superconductors, including the symmetry-breaking states of $\nu=1$. (Note that the above scheme of obtaining $\Delta G$ breaks down for the case of $\nu=1$, likely due to more complicated tunneling mechanisms and wavefunctions.) Theoretical studies on this novel type of chiral quasiparticle tunneling are still needed. We anticipate that further investigation on this tunneling conductance could help us understand the characteristics of chiral electron-hole hybrid states and ultimately pursue topological superconductivity in QH/SC graphene devices.


\begin{acknowledgments}

Low-temperature electronic measurements performed by M.T.W., A.W.D., and G.F. were supported by ARO Award W911NF16-1-0122 and NSF awards ECCS-1610213 and DMR-1743907.  Lithographic fabrication and characterization of the samples performed by M.T.W. and A.S. were supported by the Division of Materials Sciences and Engineering, Office of Basic Energy Sciences, U.S. Department of Energy, under Award DE-SC0002765. S.T. and M. Y. acknowledges KAKENHI (GrantNo. 38000131, 17H01138).  I.V.B. acknowledges CityU New Research Initiatives/Infrastructure Support from Central (APRC): 9610395, and the Hong Kong Research Grants Council (ECS) Project: 9048125. We thank Albert Chang, Harold Baranger and Gu Zhang for fruitful discussions about the data.

\end{acknowledgments}

\end{document}